\documentclass[10pt,a4paper,twoside]{article}
\usepackage{epsfig}
\usepackage{baltlat6}
\usepackage{array}
\usepackage{here}
\begin{document}
\ \
\vspace{0.5mm}
\setcounter{page}{1}
\vspace{8mm}

\titlehead{Baltic Astronomy, vol.\,}

\titleb{FORMATION OF NEUTRAL DISK-LIKE ZONE AROUND THE ACTIVE
HOT STARS IN SYMBIOTIC BINARIES}

\begin{authorl}
\authorb{Zuzana Carikov\'{a}}{} and
\authorb{Augustin Skopal}{}
\end{authorl}

\begin{addressl}
\addressb{}{Astronomical Institute, Slovak Academy of Sciences,\\
059 60 Tatransk\'{a} Lomnica, Slovakia}
\end{addressl}

\submitb{Received: 2011; accepted: 2011}

\begin{summary}
In this contribution we present the ionization structure in the
enhanced wind from the hot star in symbiotic binaries during
active phases.
Rotation of the hot star leads to the compression of the
outflowing material towards its equatorial plane. As a result
a neutral disk-like zone around the active hot star near the
orbital plane is created.
We modelled the compression of the wind using the wind
compression model. Further, we calculated the neutral disk-like
zone in the enhanced wind from the hot star using the
equation of the photoionization equilibrium.
The presence of such neutral disk-like zones was also suggested
on the basis of the modelling the spectral energy distribution
of symbiotic binaries.
We confront the calculated ionization structures in the enhanced
wind from the hot star with the observations. We found that the
calculated column density of the neutral hydrogen atoms in the
neutral disk-like zone and the emission measure of the ionized
part of the wind from the hot star are in a good agreement with
quantities derived from observations during active phases.
The presence of such neutral disk-like zones is transient,
being connected with the active phases of symbiotic binaries.
During quiescent phases, such neutral disk-like zones cannot be
created, because of insufficient mass loss rate of the hot star.
\end{summary}


\resthead{Formation of neutral disk-like zone around the hot star
during active phases}
{Z. Carikov\'{a}, A. Skopal}

\sectionb{1}{INTRODUCTION}

Symbiotic stars are long-period interacting binary systems,
which comprise a late type giant and a hot compact star
(most frequently a white dwarf). During quiescent phases
part of the material from the wind from the cool giant is
accreted by the hot compact star. Accretion process makes the
surface of the hot star to be very hot ($\sim 10^{5}$ K)
and luminous ($\sim 10^{2} - 10^{4}$ ${\rm L}_{\sun}$) and so
capable of ionizing the neutral surrounding material.
It gives rise to a strong nebular radiation
(e.g. Seaquist, Taylor \& Button 1984). Therefore,
we have three basic sources of the radiation: the
cool giant, the hot star and the nebula.

Modelling the spectral energy distribution of symbiotic
binaries during active phases showed two different cases.
In binaries with high orbital inclination the temperature
of the hot star rapidly decreases to $\sim$ 22\,000\,K,
while in systems with low orbital inclination the temperature
of the hot star increases to $\sim$ 165\,000\,K.
This inclination dependent behaviour can be explained by the
presence of the optically thick disk-like structure created
during active phases around the hot star near the orbital plane
of the symbiotic binary (Skopal 2005).
Modelling the broad H$\alpha$ wings showed that
during active phases the stellar wind from the hot star
is significantly enhanced to a few
$\times (10^{-7} - 10^{-6})$ ${\rm M}_{\sun} {\rm yr}^{-1}$
(Skopal 2006).

In our work we decided to test the idea if this enhanced wind
from the hot star can be responsible for creation of the
shielding disk-like structure around the hot star during active
phases. Further, we calculated the column density of the neutral
hydrogen atoms throughout the whole disk in direction of the
central star and the emission measure of the ionized wind from
the hot star and compared our results with quantities derived
from observations during active phases.
We found a good agreement.

\sectionb{2}{WIND COMPRESSION MODEL}

Wind compression model was developed by Bjorkman \& Cassinelli
(1993). This model describes the compression of the radiation
driven wind from the rotating hot star towards the equatorial
regions. The outflowing material is bent towards the equatorial
plane due to the conservation of the angular momentum. If the
streamlines of gas do not cross the equatorial plane, then we
are talking about the wind compressed zone (WCZ) model, which
was described by Ignace, Cassinelli and Bjorkman (1996).
We applied this model to the stellar wind from the hot star
in symbiotic binaries during active phases.

Density in the compressed wind follows from the mass continuity
equation as
\begin{equation}
  N_{\rm H}(r,\theta) = 
          \frac{\dot{M}}{4\pi r^2 \mu_{\rm m} m_{\rm H}
          v_{\rm r}(r)}\left(\frac{d\mu}{d\mu_{0}}\right)^{-1},
\end{equation}
where $\dot{M}$ is the mass loss rate of the hot star and
the compression of the wind is given by the geometrical
factor $d\mu/d\mu_{0}$. This factor depends on the rotational
velocity of the hot star as well as on the parameters of the
wind. For the wind velocity $v_{\rm r}(r)$ we adopted
$\beta$ - law. More details about the wind compression model
can be found in Lamers \& Cassinelli (1999).

\sectionb{3}{IONIZATION STRUCTURE IN THE WIND}

The ionization boundary is defined by the locus of points at
which ionizing photons are completely consumed along path
outward from the ionizing star. We calculated the ionization
boundary in the wind from the hot star using the equation of the
photoionization equilibrium. This equation equals the number
of ionizations per second with the number of recombinations
per second. For the simplicity, we assumed that the wind from
the hot star contains only hydrogen atoms.
We derived an equation for determining the ionization radius
(i.e. the ionization boundary in the direction given by the
polar angle $\theta$), which can formally be written as
\begin{equation}
  X = f(u,\theta,a,v_{\infty},\beta,v_{\rm rot}), 
\end{equation}
where $u$, $\theta$ are polar coordinates ($\theta=0$ at the
rotational axis), $a$, $v_{\infty}$, $\beta$ are parameters
of the wind, $v_{\rm rot}$ is the rotational velocity of
the hot star, and the parameter $X$ is given by
\begin{equation}
  X = \frac{8\pi \mu_{\rm m}^{2} m_{\rm H}^{2}}{\alpha_{\rm B}}
  R_{\ast} L_{\rm H} \left(\frac{v_{\infty}}{\dot{M}}\right)^{2},
\end{equation}
where $L_{\rm H}$ is the rate of photons from the hot star, 
capable of ionizing hydrogen (it is given by the temperature
$T_{\rm h}$ and luminosity $L_{\rm h}$ of the ionizing source)
and $\alpha_{\rm B}$ is the total hydrogenic recombination
coefficient in case B.
Most of the parameters of the hot star and its wind can be
determined from observations. However, the rotational velocities
of the hot stars in symbiotic binaries are not commonly known.
We calculated models for different rotational velocities from
100 to 350 km s$^{-1}$.
Figure 1 shows some examples of the calculated ionization
boundaries for two different rotational velocities of the hot
star, i.e. 200 and 300 km s$^{-1}$, respectively.

\begin{figure}
\begin{center}
\begin{tabular}{cc}
\resizebox{6.0cm}{!}{\includegraphics[angle=270,trim=0.3cm 3.5cm 0.3cm 4cm]{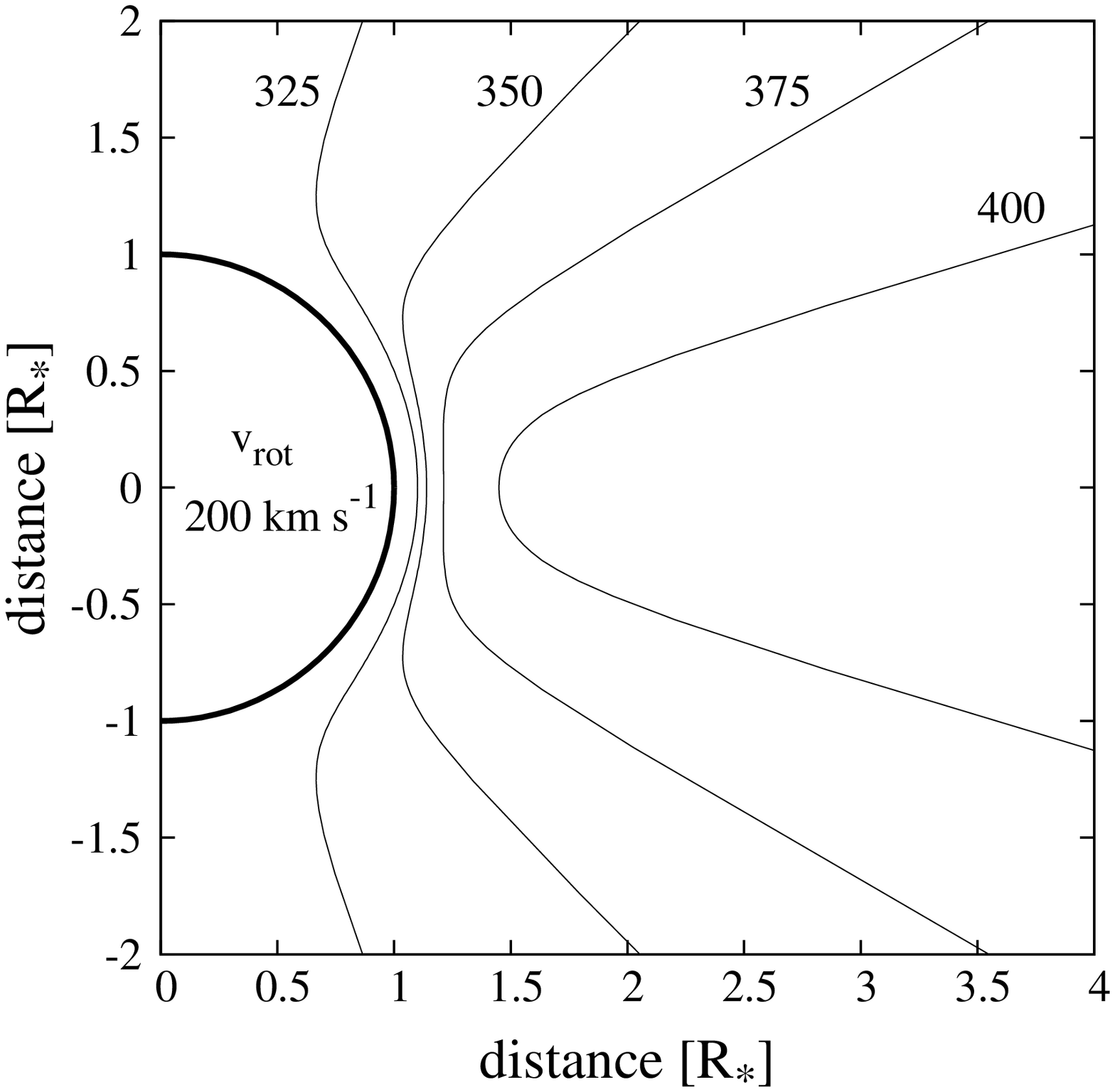}}&
\resizebox{6.0cm}{!}{\includegraphics[angle=270,trim=0.3cm 3.5cm 0.3cm 4cm]{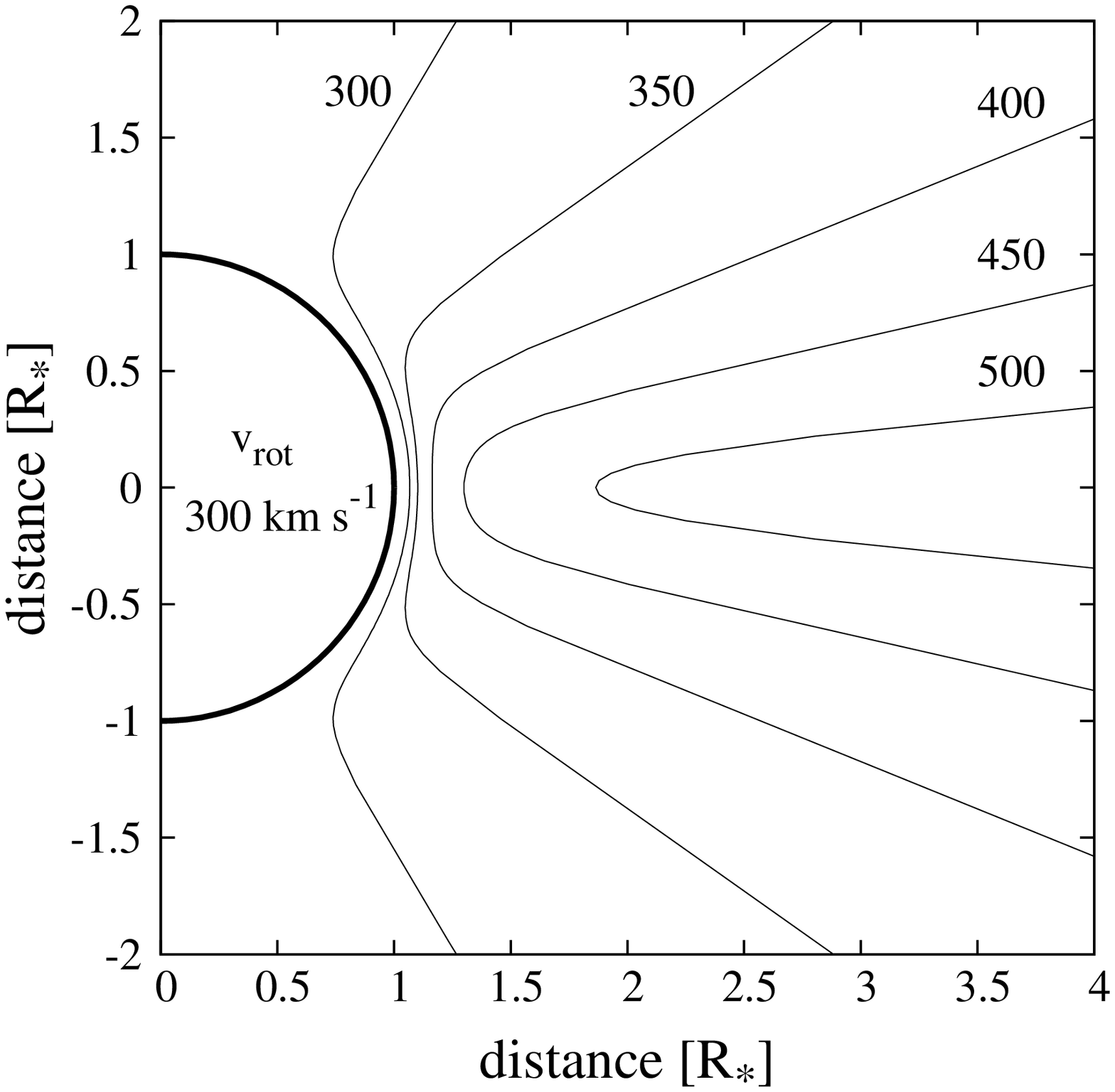}}
\end{tabular}
\end{center}
\captionb{1}{Some examples of ionization boundaries in the wind
for two different rotational velocities of the hot star:
200 and 300 km s$^{-1}$.
Individual ionization boundaries are labelled by the value of
the parameter $X$. From an ionization boundary towards the pole
of the star (y-axis) there is the ionized zone, and towards the
equatorial plane (y $=$ 0) there is the neutral zone.
Ionization structures are axially symmetric with respect to the
polar (rotational) axis of the hot star.
Distances are in units of radius of the active hot star $R_{\ast}$.}
\end{figure}

There is only a certain range of values of the
parameter $X$, for which the neutral disk-like structure can be
created near to the equatorial plane of the hot star.
For higher rotational velocities of the hot star (i.e. higher
compression of the stellar wind towards the equatorial plane)
this range is wider.
For a given rotational velocity, small values of the parameter
$X$ ($\sim 200$) correspond to ionization boundaries, which are
enclosed in the vicinity of the hot star. On the other hand,
increasing value of $X$ corresponds to moving the ionization
boundary away from the vicinity of the hot star,
as well as to decreasing the opening angle of the neutral
disk-like zone until it disappears for very high values of $X$.
However, particular values depend also on the parameters of
the wind from the hot star.

\sectionb{4}{DISCUSSION}

\subsectionb{4.1}{Quiescent phase versus active phase}

According to Eq. (3), the parameter $X$ strongly
depends on the mass loss rate of the hot star $\dot{M}$ as
\begin{equation}
  X \propto \frac{1}{\dot{M}^{2}}.
\end{equation}
Other parameters in Eq. (3) do not change
significantly between quiescent and active phase.

Typical mass loss rate of the hot star during {\em active phase}
is $\sim (10^{-7} - 10^{-6})$ ${\rm M}_{\sun} {\rm yr}^{-1}$,
while during {\em quiescent phase} it decreases to a few
$\times (10^{-8})$ ${\rm M}_{\sun} {\rm yr}^{-1}$ (Skopal 2006).
Figure 2 shows the dependence of the parameter
$X$ on the mass loss rate of the hot star
(all other parameters were fixed to typical values during
the active phase, e.g. $v_{\infty} = 2\,000$ km s$^{-1}$).
We used two different combinations of temperature $T_{\rm h}$
and luminosity $L_{\rm h}$ of the hot star during active phase,
which were derived by Sokoloski et al. (2006) for Z And
during its 2000-03 outburst.

\begin{figure}
\begin{center}
\resizebox{10.0cm}{!}{\includegraphics[angle=270]{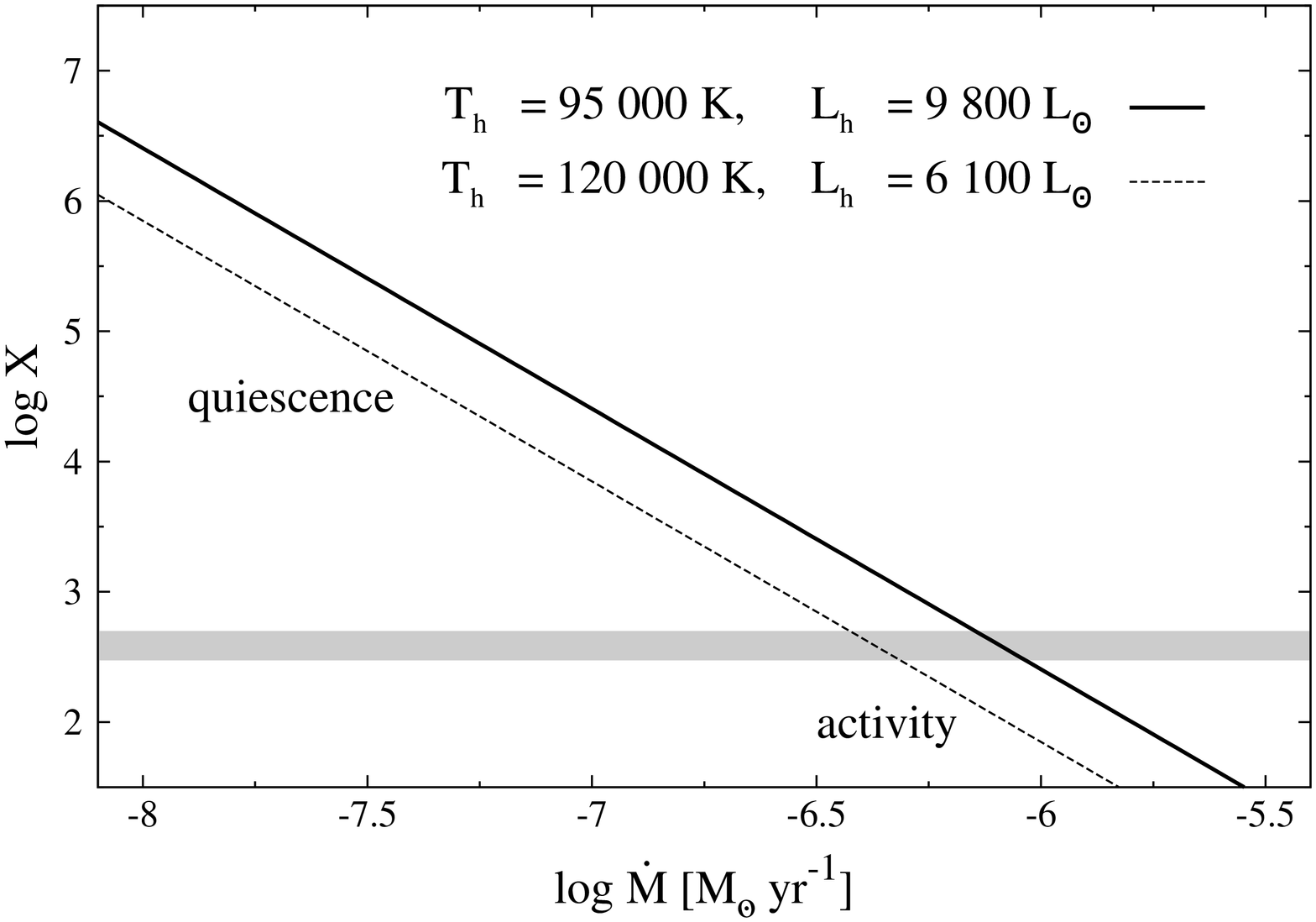}}
\end{center}
\captionb{2}{The value of the parameter $X$ as a function of
the mass loss rate ,$\dot{M}$, of the hot star for two different
combinations of the temperature $T_{\rm h}$ and luminosity
$L_{\rm h}$, derived by Sokoloski et al. (2006) for the hot
component in Z And during its 2000-03 outburst.
The shadow belt corresponds to $X=$ 300 -- 500 from Fig. 1,
which lead to creation of the neutral disk-like zone.}
\end{figure}

Our calculations (Fig. 1) showed that creation
of the neutral disk-like structure requires values of the
parameter $X$ to be of the order of hundreds.
According to Fig. 2, such the quantity of the parameter $X$
corresponds to mass loss rates of a few
$\times (10^{-7} - 10^{-6})$ ${\rm M}_{\sun} {\rm yr}^{-1}$,
which are consistent with those derived independently from the
broad H$\alpha$ wings during active phase.
Thus the neutral disk-like zone can be created during active
phases as a result of the enhanced wind from the hot star.
In contrast, during quiescent phases, the measured quantities
$\times (10^{-8} - 10^{-7})$ ${\rm M}_{\sun} {\rm yr}^{-1}$,
correspond to $X\sim 10^{4} - 10^{6}$, which are far beyond
any possibility to form a neutral disk-like zone around the
hot star.

\subsectionb{4.2}{Model versus observations}

Figure 3 shows an example of the ionization structure around
the hot star during active phases calculated for the density
distribution in the wind according the WCZ model (Section 2).
It is similar to the schematic model proposed on the basis of
multiwavelength modelling the spectral energy distribution
of symbiotic binaries (see Fig. 27 of Skopal 2005).

In this section we compare the column density of neutral
hydrogen atoms and the emission measure of the wind of our
theoretical model with those derived from observations during
active phases.

Modelling the spectral energy distribution of symbiotic binaries
showed that there is a large amount of the neutral hydrogen near
to the orbital plane of the binary. Column densities of
the neutral hydrogen atoms derived directly from the ultraviolet
spectra, run between
$\sim 10^{21}$ and a few $\times 10^{23}$ cm$^{-2}$
(Skopal 2005). This large amount of the neutral material can be
detected during active phases even near the orbital phase
$\phi \sim 0.5$ (i.e. when the hot star is in front of the cool
giant).

Further, this modelling the spectral energy distribution
showed that the emission measure of the enhanced wind from the
hot star during active phases runs between a few
$\times 10^{58}$ and a few $\times 10^{60}$ cm$^{-3}$ (Skopal 2005).

\begin{figure}
\begin{center}
\resizebox{9.0cm}{!}{\includegraphics[angle=270]{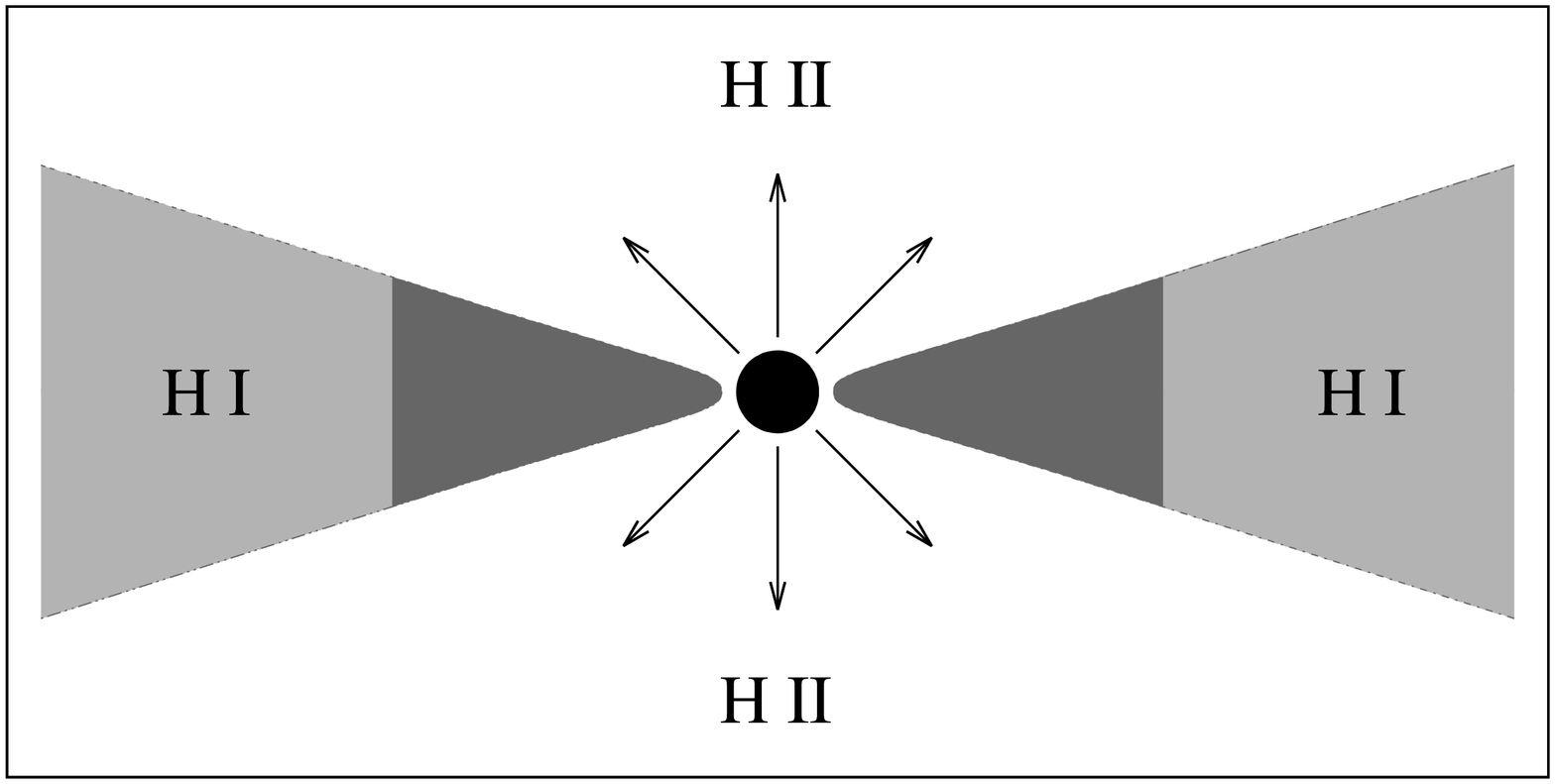}}
\end{center}
\captionb{3}{An example of the ionization structure in
the enhanced wind from the hot star during active phase
calculated for $v_{\rm rot} = 200$ km s$^{-1}$ and $X = 400$.
The hot star is denoted by the black filled circle and the
neutral zone (H I) is drawn in grey. Its shape resembles a flared
disk. The optically thick part of this neutral zone is drawn
in dark grey. The boundary between the grey and dark grey
region mimics the false photosphere. In the symbiotic binaries
with high orbital inclination we are looking at this false
photosphere, which radiates at significantly lower temperature
than the central hot star. The ionized zone (H II) is located above
and below the neutral disk-like structure and can be associated
with the hot star wind (black arrows).}
\end{figure}

\begin{figure}
\begin{center}
\begin{tabular}{cc}
\resizebox{6.0cm}{!}{\includegraphics[angle=270,trim=0.3cm 3.5cm 0.3cm 4cm]{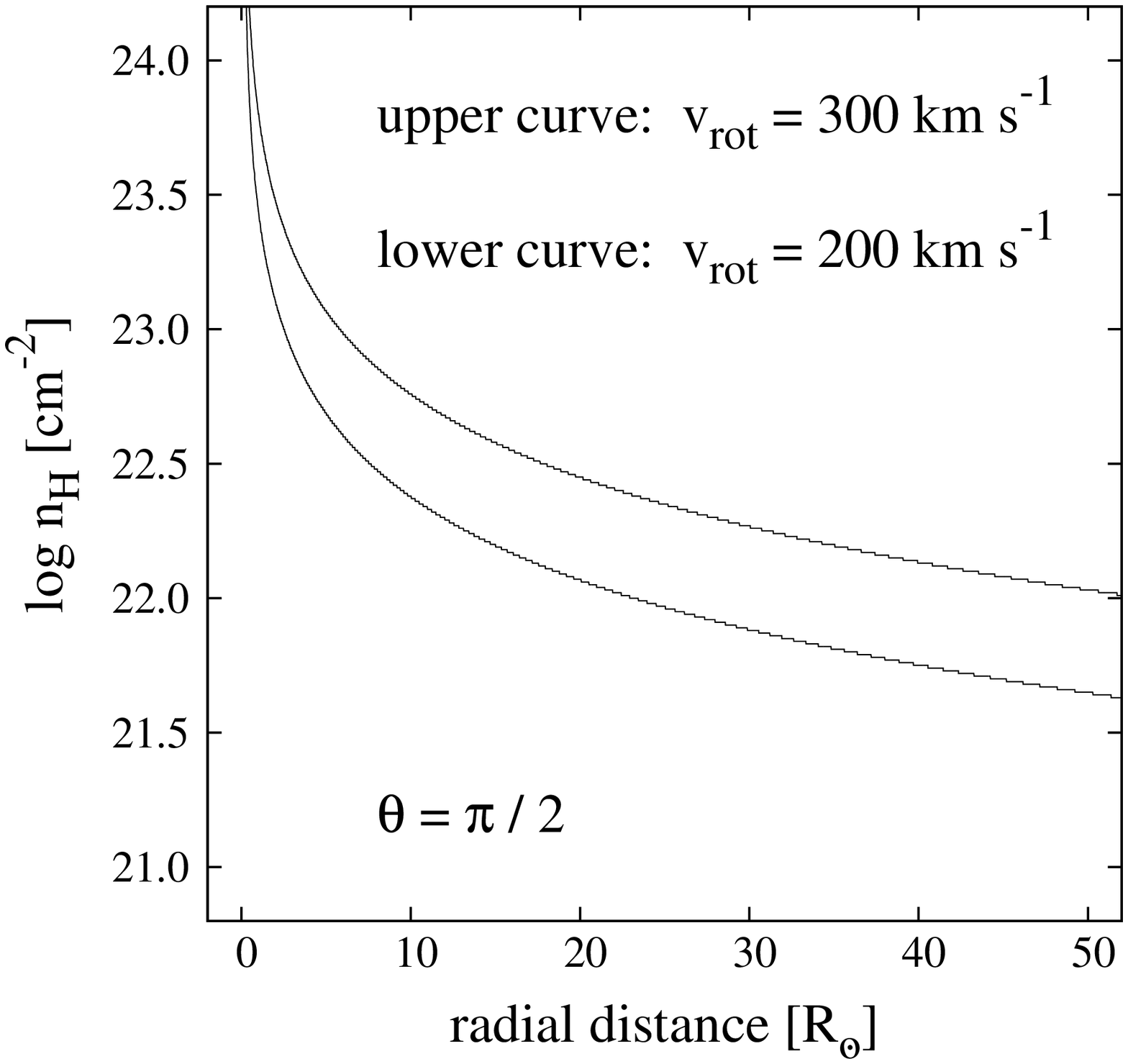}}
\resizebox{6.0cm}{!}{\includegraphics[angle=270,trim=0.3cm 3.5cm 0.3cm 4cm]{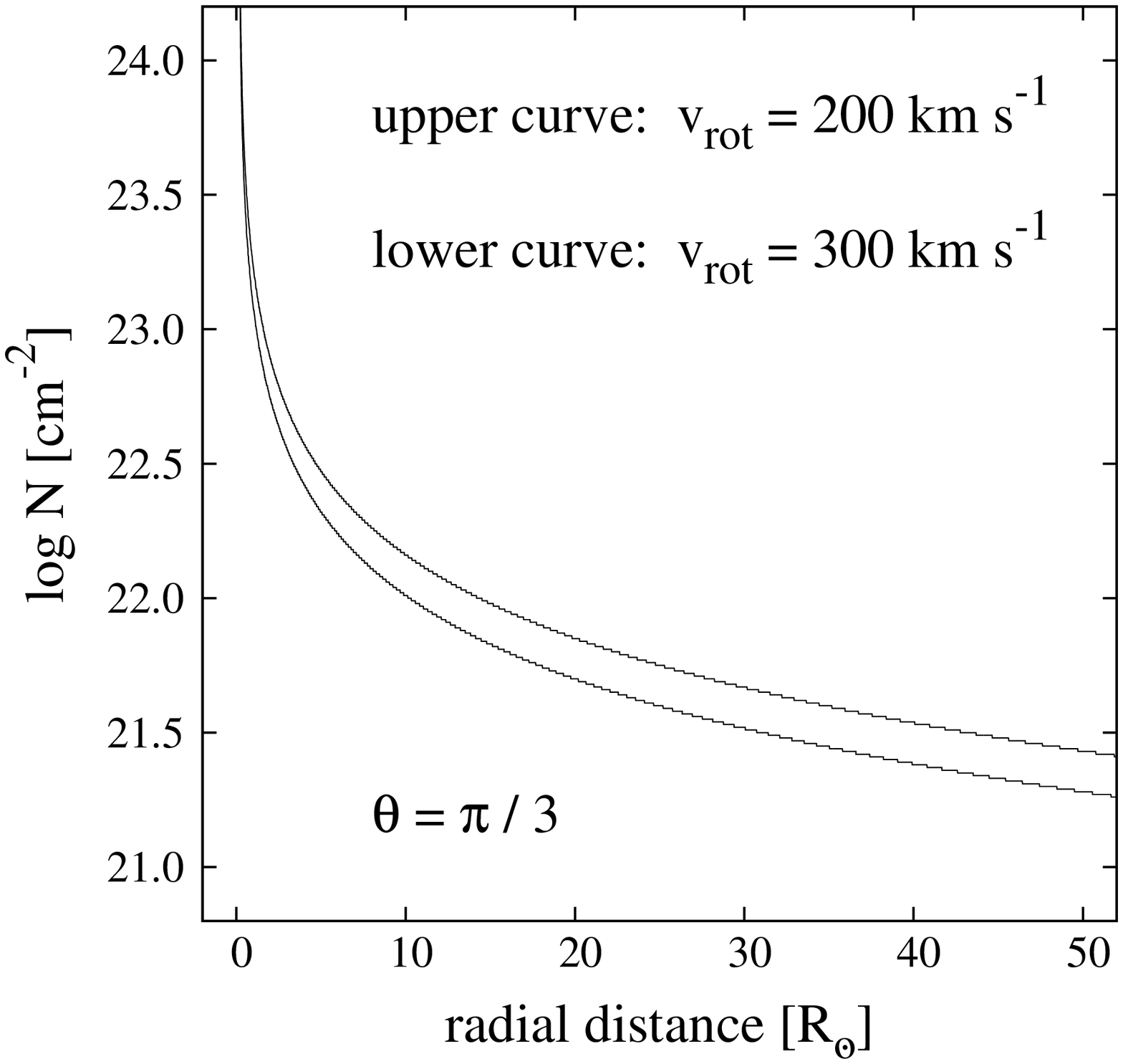}}
\end{tabular}
\end{center}
\captionb{4}{Examples of the calculated column densities
$N (\theta)$ of the neutral hydrogen atoms as a function of the
radial distance from the centre of the hot star for two different
rotational velocities, 200 and 300 km s$^{-1}$.
The angle between the polar axes of the hot star and the line
of sight was $\theta = \pi/2$ (equatorial plane) for the left
panel and $\theta = \pi/3$ for the right one.
For this calculations we assumed that the mass loss rate of
the hot star is $10^{-6}$ ${\rm M}_{\sun} {\rm yr}^{-1}$ and
the radius of the hot star is 0.18 ${\rm R}_{\sun}$.
}
\end{figure}

\begin{figure}
\begin{center}
\resizebox{6.0cm}{!}{\includegraphics[angle=270,trim=0.3cm 3.5cm 0.3cm 4cm]{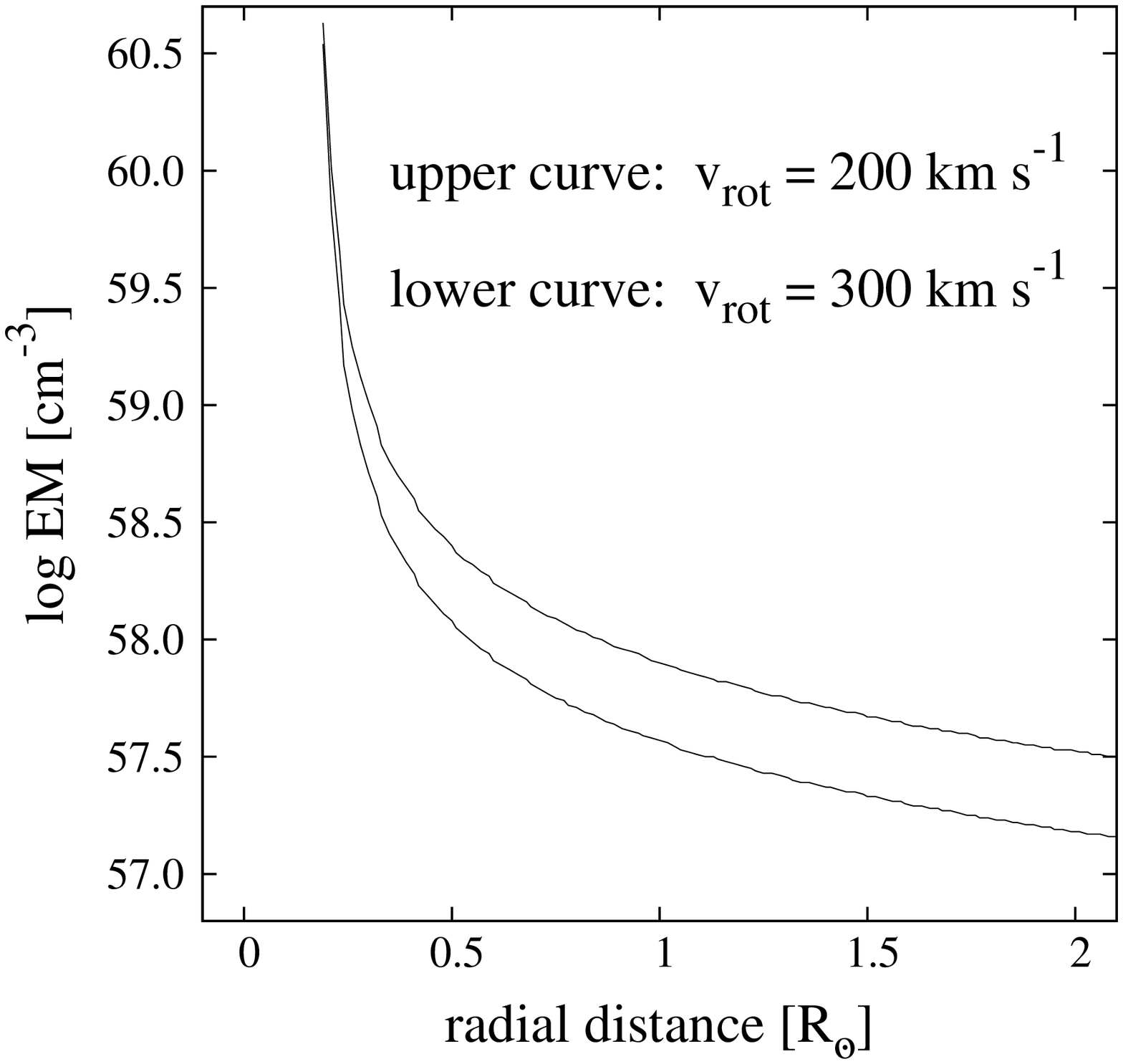}}
\end{center}
\captionb{5}{Two examples of the calculated emission measures
$EM$ of the wind from the hot star as a function of the
radial distance from the centre of the hot star.
The emission measures were calculated for the hot star with
the mass loss rate $10^{-6}$ ${\rm M}_{\sun} {\rm yr}^{-1}$.
Further, we assumed that the radius of the hot star is
0.18 ${\rm R}_{\sun}$ and the boundary between neutral
and ionized zone is given by the polar angle $\theta = \pi/3$.}
\end{figure}

Theoretical column density of the neutral hydrogen $N(\theta)$
in the direction $\theta$ throughout the neutral zone was
calculated as
\begin{equation}
  N(\theta) = \int_{r} N_{\rm H} (r,\theta) \,{\rm d}r,
\end{equation}
where $N_{\rm H} (r,\theta)$ is the density of the neutral
hydrogen given by Eq. (1).
Figure\,\,4 shows examples of the calculated column densities
of the neutral hydrogen as a function of the radial distance
from the centre of the hot star for two different rotational
velocities of the hot star, 200 and 300 km s$^{-1}$.
Since the radius of the false photosphere can be in the order
of tenths of solar radii, we can easily see that the
calculated values of the column densities of the neutral
hydrogen are in a good agreement with those derived from
observations.

Theoretical emission measure $EM$ of the ionized wind
from the active hot star was calculated throughout the ionized
zone as
\begin{equation}
  EM = \int_{V} N_{\rm H}^{2} (r,\theta) \,{\rm d}V,
\end{equation}
where $N_{\rm H} (r,\theta)$ is the density of the ionized
hydrogen given by Eq. (1).
In spherical coordinates the volume element ${\rm d} V$ is
\begin{equation}
  {\rm d} V = r^{2} \sin \theta \,{\rm d}r\,{\rm d}\theta\,{\rm d}\phi.
\end{equation}
The density distribution in the WCZ model is azimuthally
symmetric. Therefore Eq. (6) can be simplified to
\begin{equation}
  EM = 2\pi \int_{r} \int_{\theta} N_{\rm H}^{2} (r,\theta)\,r^{2}\sin \theta\,{\rm d}r\,{\rm d}\theta.
\end{equation}
Figure 5 shows two examples of the calculated emission measures
as a function of the radial distance from the centre of the hot
star for two different rotational velocities, 200 and 300
km s$^{-1}$.
From Fig. 5 we can see that the calculated values of the
emission measure of the ionized wind from the hot star
are in a good agreement with the observed quantities.

\sectionb{5}{CONCLUSION}

We found that the compression of the enhanced stellar wind
in active phases from the rotating hot star towards
equatorial regions can lead to the creation of the neutral
disk-like structure around the hot star near the orbital
plane.
Figure\,\,1 shows some examples of ionization boundaries in the
wind from the active hot star in symbiotic binaries calculated
for the density distribution within the WCZ model.
In symbiotic binaries with high orbital inclination
we are viewing through this neutral disk-like zone.

Theoretical ionization structure (e.g. Fig. 3), which we
calculated using the WCZ model, is consistent with that derived
on the basis of the multiwavelength modelling the spectral
energy distribution of symbiotic binaries during active phases
(see Fig. 27 of Skopal 2005).
Also the calculated column density of the neutral hydrogen atoms
throughout the neutral zone and the emission measure of the
ionized region are in a good agreement with the quantities
derived from observations during active phase.
The large value of the mass loss rate of the hot star during
active phases, a few
$\times (10^{-7} - 10^{-6})$ ${\rm M}_{\sun} {\rm yr}^{-1}$,
derived from the broad H$\alpha$ wings (Skopal 2006) support
the plausibility of the used model.

We explained that due to a low mass loss rate from
the hot star, no neutral disk-like structure can be created
during quiescent phases.

\thanks{This research was supported by a grant of the Slovak
Academy of Sciences, VEGA No. 2/0038/10.}

\References

\refb Bjorkman, J. E., Cassinelli, J. P. 1993, ApJ, 409, 429
\refb Ignace, R., Cassinelli, J. P., Bjorkman, J. E. 1996, ApJ,
      459, 671
\refb Lamers, H. J. G. L. M., Cassinelli, J. P. 1999,
      Introduction to stellar winds, Cambridge University Press
\refb Seaquist, E. R., Taylor, A. R., \& Button, S. 1984, ApJ,
      284, 202		 
\refb Skopal, A. 2005, A\&A, 440, 995
\refb Skopal, A. 2006, A\&A, 457, 1003
\refb Sokoloski, J. L., Kenyon, S. J., Espey, B. R. et al. 
      2006, ApJ, 636, 1002		

\end{document}